\documentclass[conference]{IEEEtran}
\IEEEoverridecommandlockouts
\usepackage{cite}
\usepackage{amsmath,amssymb,amsfonts}
\usepackage{algorithmic}
\usepackage{graphicx}
\usepackage{textcomp}
\usepackage{xcolor}
\usepackage{booktabs}
\usepackage{tabularx}
\def\BibTeX{{\rm B\kern-.05em{\sc i\kern-.025em b}\kern-.08em
    T\kern-.1667em\lower.7ex\hbox{E}\kern-.125emX}}
\begin{document}

\title{Lightly Weighted Automatic Audio Parameter Extraction for the Quality Assessment of Consensus Auditory-Perceptual Evaluation of Voice \\
%{\footnotesize \textsuperscript{*}Note: Sub-titles are not captured in Xplore and should not be used}
\thanks{Corresponding author: Ching-Ting Tan (tanct5222@ntu.edu.tw). Yi-Heng Lin and Wen-Hsuan Tseng contributed equally. Li-Chin Chen and Ching-Ting Tan contributed equally.}
}

\author{\IEEEauthorblockN{Yi-Heng Lin}
\IEEEauthorblockA{
\textit{Department of Electrical Engineering,}\\
\textit{National Taiwan University} \\
\textit{Research Center for Information Technology Innovation,}\\
\textit{Academia Sinica}\\
Taipei, Taiwan \\
b08901161@ntu.edu.tw}
\and
\IEEEauthorblockN{Wen-Hsuan Tseng}
\IEEEauthorblockA{
\textit{Department of Otolaryngology,}\\
\textit{National Taiwan University Hospital}\\
\textit{Graduate Institute of Clinical Medicine,} \\
\textit{National Taiwan University College of Medicine}\\
Taipei, Taiwan \\
100881@ntuh.gov.tw}
%ORCID: 0000-0002-0179-4637
\and
\IEEEauthorblockN{Li-Chin Chen}
\IEEEauthorblockA{\textit{Research Center for Information Technology Innovation,} \\
\textit{Academia Sinica}\\
Taipei, Taiwan \\
li.chin@citi.sinica.edu.tw}
\and
\IEEEauthorblockN{Ching-Ting Tan}
\IEEEauthorblockA{
\textit{Department of Otolaryngology,}\\
\textit{National Taiwan University College of Medicine}\\
\textit{Intelligent Healthcare Innovation Center,}\\
\textit{National Taiwan University Hospital Hsin-Chu Branch}\\
Taipei, Taiwan \\
tanct5222@ntu.edu.tw}
%ORCID: 0000-0001-8317-2235}

\and
\IEEEauthorblockN{Yu Tsao}
\IEEEauthorblockA{\textit{Research Center for Information Technology Innovation,} \\
\textit{Academia Sinica}\\
Taipei, Taiwan \\
yu.tsao@citi.sinica.edu.tw}

}

\maketitle

\begin{abstract}
The Consensus Auditory-Perceptual Evaluation of Voice is a widely employed tool in clinical voice quality assessment that is significant for streaming communication among clinical professionals and benchmarking for the determination of further treatment. Currently, because the assessment relies on experienced clinicians, it tends to be inconsistent, and thus, difficult to standardize. To address this problem, we propose to leverage lightly weighted automatic audio parameter extraction, to increase the clinical relevance, reduce the complexity, and enhance the interpretability of voice quality assessment. The proposed method utilizes age, sex, and five audio parameters: jitter, absolute jitter, shimmer, harmonic-to-noise ratio (HNR), and zero crossing. A classical machine learning approach is employed. The result reveals that our approach performs similar to state-of-the-art (SOTA) methods, and outperforms the latent representation obtained by using popular audio pre-trained models. This approach provide insights into the feasibility of different feature extraction approaches for voice evaluation. Audio parameters such as jitter and the HNR are proven to be suitable for characterizing voice quality attributes, such as roughness and strain. Conversely, pre-trained models exhibit limitations in effectively addressing noise-related scorings. This study contributes toward more comprehensive and precise voice quality evaluations, achieved by a comprehensively exploring diverse assessment methodologies. 

%contemporary studies have concentrated on automating the evaluation process through computer assistance . Prior research has identified certain audio parameters that contribute to assessing the severity of voice disorders. Concurrently, the surge of deep learning enables powerful pre-trained mode of learning and extracting the representation of voice. In our study, we take benefit of the two feature extraction methods and compare their performance against the state-of-the-art (SOTA) results. The approach based on extracting audio parameters attains results comparable to the SOTA while offering reduced complexity and greater clinical relevance. In contrast, the utilization of pre-trained models proves inadequate in yielding satisfactory outcomes.
\end{abstract}

%% 8/14 abstract 的結論可能要改嗎


% 要點整理：
% 1. 雖然縮寫已經在abstract寫過了，但是正文裡面還是要再寫一次 done 
% 2. 摘要裡面不要有references  done
% 3. \cite{} 可以吃好幾個參數  done 
% 4. 直接空一行，PDF就會換成下一段了  done 

\section{Introduction}

Clinical voice quality evaluation describes the severity of distinct vocal irregularities, encompassing attributes such as roughness or breathiness. This process aids in determining the necessity for additional treatment or diagnostic procedures. However, the evaluation must be performed by clinicians with several years of experience, and the variability in outcomes complicates in the subsequent assessment of therapeutic regimens \cite{Voice_quality}. To minimize these inconsistencies, developing an automated evaluation tool can be beneficial. Several studies have been conducted to automate the evaluation process by using computer-aided methodologies, including classical machine learning techniques \cite{scatter_wavelet,GRBAS_1,GRBAS_2,GRBAS_3,GRBAS_6} or deep neural networks \cite{GRBAS_5}.

The extraction processes implemented in these studies have been considerably different. The intricate nature of the raw audio signals makes producing convincing scores by computers challenging. The approach chosen for feature extraction significantly influences the predictive outcomes of machine learning algorithms. Currently, the best performing method utilizes scatter wavelet transform for feature extraction \cite{scatter_wavelet}. The scatter wavelet transform is a multi-layer signal decomposition technique that effectively functions as a convolutional network with wavelet filters. However, its extracted features are ambiguous, and difficult to correspond to audio characteristics and cannot provide insights into existing knowledge. Moreover, the data size used in previous studies ranged from 200 to 300 \cite{scatter_wavelet, PVQD, GRBAS_6}, which is considered insufficient for training deep neural networks. To address this limitation, pre-trained models had been known as improving a neural network in one domain using more readily obtainable data, and subsequently harnessing the accrued knowledge to overcome the limitations imposed by the dataset size \cite{SSL_survey}. %Therefore, we aim to leverage well-known pre-trained models to transcend data size constraints. 

We propose a lightly weighted feature extraction approach based on well-established audio parameters and classical machine learning techniques. This approach provides insights that correspond to audio characteristics and interpretability during machine learning. In addition, since the pre-training models are known to overcome the limitations of small sample sizes, we examine their efficiency by leveraging well-known pre-trained models during feature extraction. 
% 回應 data size 的問題， 目前work的 data size 都很小，呼應我們為什麼要使用 pre-trianed model. 

%To be more specific, \cite{scatter_wavelet} applied Perceptual Voice Qualities Database \cite{PVQD}, which comprises merely 296 audio recordings. \cite{GRBAS_6} used the well-known Massachusetts Eye and Ear (MEEI) Dataset, which includes 53 normal and 173 pathological records. These existing open sources dataset suffer from limited data size, not to mention the self-organized dataset used in other result.} 

%the resulting information is ambiguous. It is hard to map the extracted information in the feature with audio characteristic. %Moreover, the transformation technique translates the input data into the time-frequency domain, potentially leading to the loss of crucial information during the transformation process.

%\cite{scatter_wavelet} reported that the calculated coefficients indicate the existence of frequency modulations within the signals.

\subsection{Scale of voice evaluation}
Consensus Auditory-Perceptual Evaluation of Voice (CAPE-V) and Grade, Roughness, Breathiness, Asthenia, Strain (GRBAS) are the most commonly used scales for voice evaluation. CAPE-V includes assessments of (a) overall severity, (b) roughness, (c) breathiness, (d) strain, (e) pitch, and (f) loudness. It also considers sex, age and referent culture during strain, pitch and loudness scoring. CAPE-V rates each sample with respect to its deviance from normal voice on a 0-100 scale. GRBAS involves the assessment of (a) overall grade, (b) roughness, (c) breathiness, (d) aesthenia, and (e) strain \cite{GRBAS_intro} on a 4-level scale. Although GRBAS is widely used, \cite{GRBAS_CAPEV} it has exhibited reduced sensitivity compared with CAPE-V when assessing small variations in voice quality. Therefore, we aim to automate voice evaluation using the CAPE-V scale.

\subsection{Feature extraction of voice}
In previous studies, various approaches for feature extraction have been explored. Generally, they can be categorized into three types: audio parameter extraction, audio transformation, and pre-trained model feature extraction. Established parameters such as the fundamental frequency, jitter, shimmer, and harmonic-to-noise ratio (HNR) have been employed for voice evaluation in several works \cite{GRBAS_1,GRBAS_3,GRBAS_5, Disorder_1}. The methods in the aforementioned studies focus on specific properties of audio samples. For instance, jitter characterizes the variation in the fundamental frequency over a short interval, whereas shimmer characterizes the amplitude variation. These parameters have been reported to be useful in voice evaluation and clinical usage.

Another type of feature extraction is based on transformation techniques. In \cite{scatter_wavelet}, a scatter wavelet transform was utilized to derive features from input audio, whereas in \cite{GRBAS_2,GRBAS_6}, Mel frequency cepstral coefficients (MFCC) or modulation spectra were applied for feature extraction. 
Unlike the computation of audio parameters, such as jitter or shimmer, transformation-based extraction does not focus on specific properties. Instead, it transforms the input audio data into another domain to highlight the important part. This method is not as specific as a calculated parameter; however, it is more general and retains more information from raw waveforms.

The final type involves deep pre-trained models. With the growth of deep learning, many powerful pre-trained models, such as the vision transformer, Wav2vec2.0 and WavLM  \cite{transformer}, have performed well in audio processing. These models excel in tasks beyond their original domains, thereby acting as excellent feature extractors. The learned latent features can be feasibly used for performing multiple tasks, thereby facilitating their deployment in cross-domain tasks. This capacity is invaluable in domains with prevailing data scarcity. Moreover, the pre-trained phase enables the model to retain valuable voice information, reducing the loss of crucial information during feature extraction. Previous studies have reported that this method is successful in biomedical research \cite{Covid_pretrained}.
% \cite{transformer} stated that pretrained model is capable of learning audio knowledge, such as jitter, shimmer. These studies encourage us to utilize pretrained model technique in voice evaluation. In our studies, we use several different transformer based pre-trained model to extract feature for further analysis. 

\section{Methods}
We apply a lightly weighted audio parameter extraction approach, with the classical machine learning methodology used in Cape-V prediction. Additionally, we utilize several renowned pre-trained models as feature extractors, subsequently fine-tuning them for enhanced Cape-V prediction. Our findings are juxtaposed with the presently acknowledged State-of-the-Art (SOTA) \cite{scatter_wavelet} benchmarks.

\subsection{Lightly weighted audio parameter extraction}
By incorporating patient information such as age and sex, we utilize well-established audio parameters, namely jitter, shimmer, HNR, and zero crossing \cite{Audio_param}, to predict the CAPE-V scores of voice samples. Jitter represents the absence of precise control over vocal cord vibrations, leading to elevated jitter values in patients with voice disorders. Notably, there are different methods in determination of jitter. Commonly used methods for determining jitter involve local jitter and local absolute jitter. The local absolute jitter method calculates the average absolute difference between two consecutive periods, whereas the local jitter method divides the local absolute jitter by the average period. In this study, we employ both. Shimmer represents reduced glottal resistance and the presence of vocal cord lesions, resulting in noise emission problems and breathiness. The HNR characterizes the ratio of the harmonic sound to the noise component, which serves as an indicator of speech efficiency. Thus, a low HNR is expected to produce a high CAPE-V score. Zero-crossing reflects the rate of sign changes in the signal, which can be construed as an indicator of the noise level of the signal \cite{ZCR}. These features help in simulating GRBAS scores \cite{GRBAS_1,GRBAS_3}, and strongly correlate with audio characteristics and voice disorders. 

Our application is designed as a regression task aimed at predicting CAPE-V scores. In the audio parameter method, the extracted feature is one-dimensional, and parameter extraction operates independently for each feature. Classical machine learning methods had demonstrated superior capability in small sample size and efficacy in identifying specific critical thresholds within calculated parameters \cite{APM59491}. Therefore, we opted for classical machine learning methods based on diverse mechanisms, including Random Forest (RF) algorithm, Support Vector Machine (SVM) and k-Nearest Neighbors (KNN). This work was implemented with Python version 3.8, the audio parameters were extracted by using Surfboard 0.2.0, and the model training were implemented through scikit-learn 1.3.0 and Pytorch 2.0.1.%Implementation of these methods is carried out using the scikit-learn module \cite{scikit-learn}. \\

\subsection{Advanced pre-trained models} %%% 要補所有用到的pretrain model 簡介
We adopt well-known audio pre-trained models: Wav2vec2, Hubert, WavLM, and Whisper. Wav2vec2 \cite{wav2vec} and Hubert \cite{hubert} are transformer-based pre-trained models \cite{transformer} capable of learning powerful representations from speech audio. Notably, they are efficient in extracting features from diverse vocal expressions such as coughs, breaths, and speech, and contributed to the detection of COVID-19 infection \cite{Covid_pretrained}. WavLM is trained on masked speech, focusing on the prediction of both the speaker and spoken content \cite{WavLm}; Whisper \cite{Whisper} is a framework that involves large-scale semi-supervised pre-training, and focuses on multilingual and multitask training. Both WavLM and Whisper intend to learn a more universal representation instead of limiting themselves to automatic speech recognition (ASR). %well not only in the ASR  but also in other domains closely tied to audio characteristics. \cite{Whisper} }

%Powerful transformer based pre-trained models such as wav2vec2 and hubert are capable of learning latent representations of audio waveform, which is reported useful in Automatic Speech Recognition(ASR) \cite{wav2vec}. It is worth noting that \cite{transformer} highlights the capability of transformer model in learning audio characteristics, showing that their utility not only in ASR but also in other domains. Additionally, the learned representation is also reported helpful in detection of COVID-19 infection \cite{Covid_pretrained}. In our work, we try to take advantage of the powerful pre-trained models to extract features. We have applied four different model : wav2vec2-base, Hubert-base, WaveLM and Whisper. % 這個不用寫 For the wav2vec2 and Hubert models, we utilize the versions and load the pre-trained weight provided by PyTorch. For WaveLM and Whisper, we employ the one available from Hugging Face.  %% Cite pytorch and hugging face. Each model extract 768 dimensional SSL features from input waveform. 

After the feature extraction process from the pre-trained models is completed, the latent representations are fine-tuned to obtain the CAPE-V score. In Wav2vec2, Hubert, and WavLM, the fine-tuned layers encompass convolution-based regressors tailored to accommodate two-dimensional features. For Whisper, a fully connected (FC) layer-based regressor is employed, because of the relatively smaller dimensions of the extracted features compared with those of the three models utilizing convolutional neural networks (CNNs). The CNN-based regressor comprises two one-dimensional convolution layers. The output of each layer subsequently passes through a batch normalization layer and the rectified linear unit (ReLU) activation function. The convolution layers are followed by three FC layers with 512, 64, and 1 output neurons respectively. Dropout layers are incorporated between these FC layers, with a probability of 0.2 to avoid overfitting, and the ReLU activation function is applied. The FC layer-based regressor for Whisper comprises three FC layers, with 256, 16, and 1 output neurons respectively. The dropout layers and ReLU function are applied in the same manner as in the CNN based framework.

% Testing 沒有 augment 要註解好
To appropriately train the fine-tuned layers, data augmentation is performed to increase the size of the training data. Audio samples with seven different types of noise are mixed, encompassing five colored noises (white, blue, violet, brown, and pink) and  two background noises featuring human speech. For each noise type, a different weight pair is employed for mixing, thereby yielding 14 augmented instances per audio sample. 
%All these extracted feature are then passed to neural network described in the following section.

\subsection{Dataset}
The audio samples used in this study are retrieved from the PVQD dataset \cite{PVQD}. It contains 296  audio recordings of sustained /a/ and /i/ vowels and sentence reading. Signals are recorded at a sampling frequency of 44.1 kHz. The voice quality of the audio samples is assessed by experienced raters based on the CAPE-V and GRBAS scales. Supplementary details, including the age and sex of the speaker, are provided. We focus on the /a/ phoneme and exclude records with corrupt files and incomplete information. The resultant dataset includes 283 patients. The average values provided by multiple raters are set as the ground truth, thereby consolidating a robust foundation for the analysis.

To reduce complexity, we downsample the audio signals to a frequency of 8 kHz, and compute the mean length across all audio recordings. For samples exceeding the average duration, additional portions are discarded. Conversely, for recordings shorter than the average length, we replicate the audio sample to double the original duration and discard the extra portions after length augmentation. This allows us to ensure uniformity in the lengths of all audio samples.

\subsection{Training strategies}
\begin{figure}
\centerline{\includegraphics[width=\columnwidth]{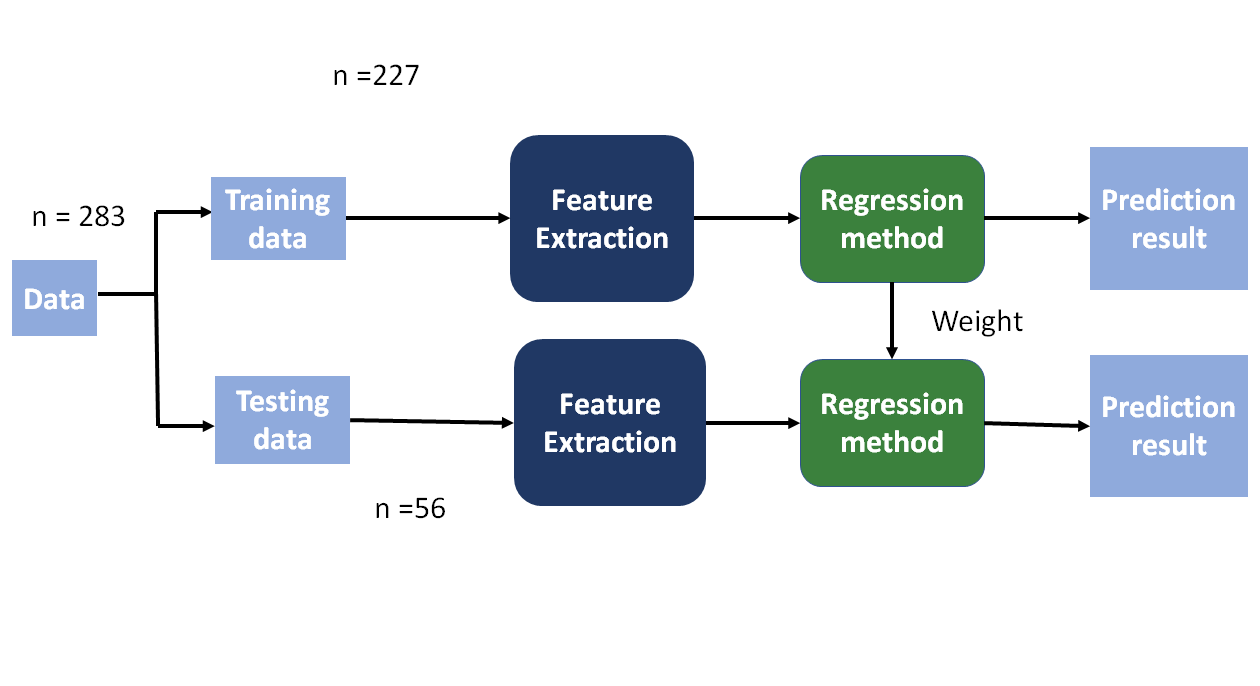}}
\caption{Experiment process of the audio parameter method.}
\label{fig : exp_diagram}
\end{figure}

\begin{figure}
\centerline{\includegraphics[width=\columnwidth]{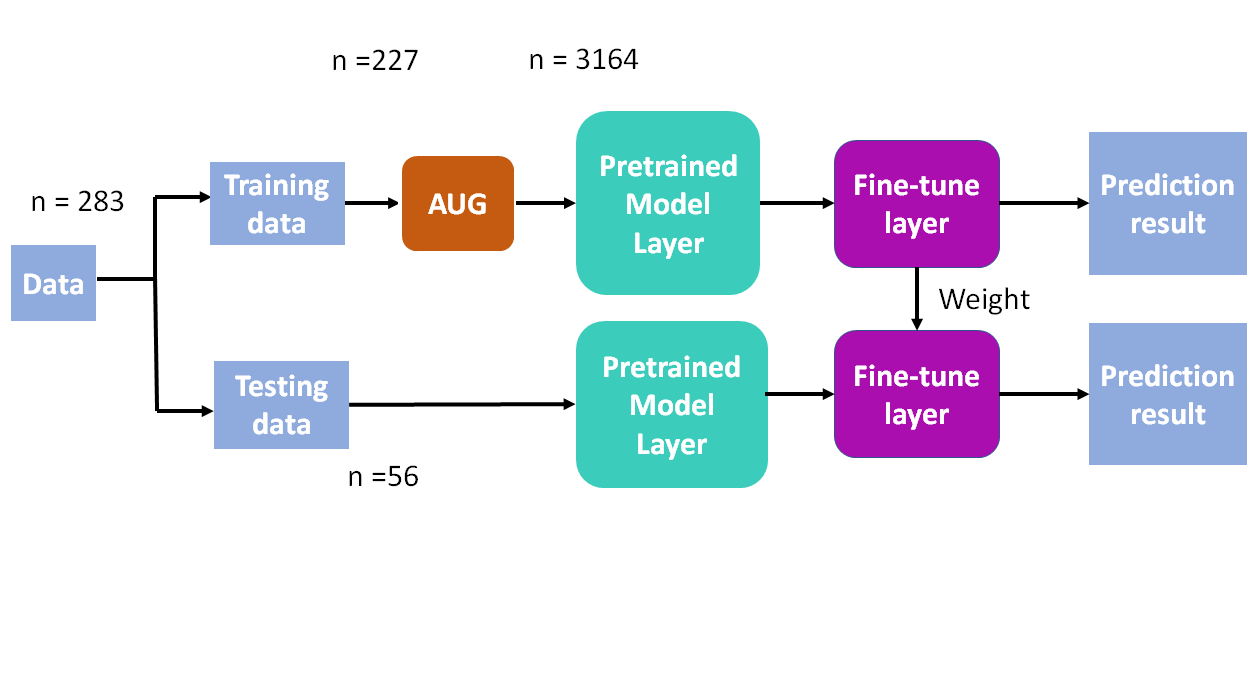}}
\caption{Experiment process of the pre-trained model method.}
\label{fig : exp_diagram1}
\end{figure}

In our experiment, the audio samples are first divided into training and testing using a balanced sampling technique, which ensures that each dataset includes samples representing a range of both high and low scoring values. This guarantees a comprehensive representation of the data distribution and enhances the reliability and validity of the evaluation of our model.\\ 
\indent During the training phase of our regression model using machine learning methods, we utilize the grid search technique for hyperparameter tuning. For the neural network model, hyperparameter tuning is performed manually. The final prediction results are compared to the Cape-V scores. In this phase, the root-mean-square error (RMSE) is utilized as the primary evaluation metric to measure the difference between predicted score and the ground truth. Pearson's correlation and the feature importance of the best performing approach are calculated to further analyze the prediction results. In addition, scatter plots are presented to visualize the difference between the ground truth and the predicted results. The experimental procedure is illustrated in Figs.~\ref{fig : exp_diagram} and \ref{fig : exp_diagram1}.  Notably, data augmentation is exclusively introduced to train the fine-tuned layers, whereas the testing phase employs un-augmented data. 
% for neural network training separately to avoid information leakage. 

%%%% 圖的邊線可能要trim一下

\section{EXPERIMENTAL RESULTS}

%%% 因為這屬於對於資料本身的探索，所以放前面
\begin{table}[htbp]
  \centering
  \caption{Correlation value of the CAPE-V score and the audio parameters}
  \label{tab:correlation}
  \begin{tabular}{c c c c c c}
    \midrule
    Method & jitter & shimmer & HNR & zero-crossing \\
    \midrule
    Severity    & 0.62 & 0.60 & -0.68 & 0.39 \\
    Breathiness & 0.54 & 0.54 & -0.57 & 0.34 \\
    Pitch       & 0.61 & 0.47 & -0.58 & 0.40 \\
    Loudness    & 0.61 & 0.54 & -0.61 & 0.40 \\
    Roughness   & 0.47 & 0.51 & -0.61 & 0.28 \\
    Strain      & 0.64 & 0.53 & -0.65 & 0.47 \\
    \bottomrule
  \end{tabular}
\end{table}

\begin{table}[htbp]
  \centering
  \caption{RMSE Results with PVQD Scores using audio parameters as features}
  \label{tab:prediction_comparison}
  \begin{tabular}{c c c c c }
    \midrule
    Method & SVM & KNN & RF & SOTA \\
    \midrule
    Severity    & 18.50 & 18.50 & 15.88 & 15.01\\
    Breathiness & 16.81 & 16.90 & 15.47 & 12.88 \\
    Pitch       & 15.56 & 15.77 & 15.10 & 14.05 \\
    Loudness    & 18.36 & 18.60 & 16.21 &  14.69\\
    Roughness   & 14.45 & 14.74 & 14.72 & 14.76 \\
    Strain      & 15.38 & 15.15 & 13.99 & 14.91 \\
    \midrule
    Avg. & 16.51 & 16.61& 15.22 & 14.38 \\
    \bottomrule
  \end{tabular}
\end{table}

\begin{table}[htbp]
  \centering
  \caption{RMSE results based on advanced pre-trained models}
  \label{tab:prediction_comparison_NN}
  \begin{tabular}{c c c c c c}
    \midrule
    Method & Wav2vec2 &  Hubert & WavLM & Whisper & RF \\
    \midrule
    Severity    & 17.97 & 20.20 & 26.92 & 16.83 & 15.88\\
    Breathiness & 14.12 & 16.22 & 20.26 & 14.85 & 15.47\\
    Pitch       & 13.25 & 15.82 & 15.82 & 13.26 & 15.10\\
    Loudness    & 18.92 & 18.36 & 16.33 & 14.26 & 16.21\\
    Roughness   & 18.66 & 19.27 & 23.33 & 18.80 & 14.72\\
    Strain      & 19.66 & 18.99 & 18.72 & 16.05 & 13.99\\
    \midrule    
    Avg. & 17.09  & 18.14 & 20.23 & 15.67 & 15.22 \\
    \bottomrule
  \end{tabular}
\end{table}

Table~\ref{tab:correlation} presents the correlation between the used audio parameters and the CAPE-V score. Jitter, shimmer, and the HNR yield correlation values of approximately 0.50-0.65, which strongly confirms the reliability of the voice evaluation result. Table~\ref{tab:prediction_comparison} lists the RMSE results of the lightly weighted audio parameter extraction. The calculated RMSE values are within 14-18. The proposed method outperforms the SOTA in terms of the roughness and strain attributes. For other attributes, our method performs similar to the SOTA approach. Table~\ref{tab:prediction_comparison_NN} presents the RMSE results of feature extraction based on the pre-trained models. The resultant RMSE values are within 13-20. The proposed method outperforms all pre-trained feature extractors in terms of the average RMSE and correlation value. %outperform in some attributes while not working well in others. \\

\begin{figure}
\centerline{\includegraphics[width=\columnwidth]{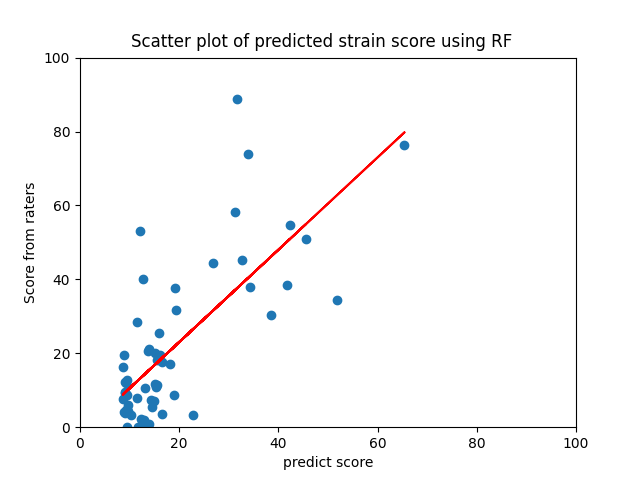}}
\caption{Scatter plot of the strain score using RF. The X-axis represents the predicted score value, and the Y-axis represents the value given by raters.}
\label{fig : strain}
\end{figure}

\begin{figure}
\centerline{\includegraphics[width=\columnwidth]{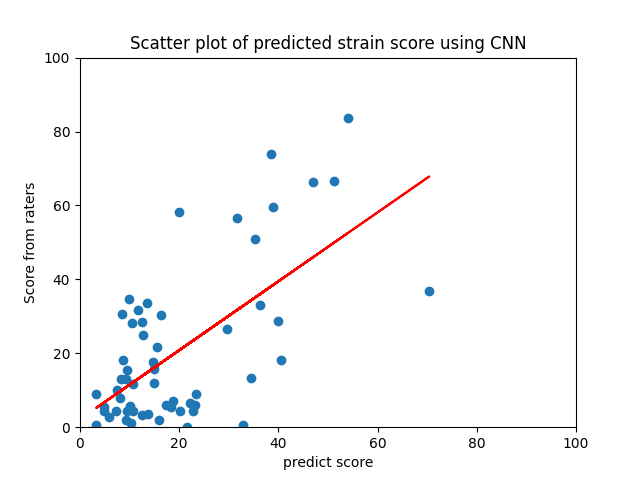}}
\caption{Scatter plot of the strain score using CNN. X-axis represents the predicted score value, and the Y-axis represents the value given by raters.}
\label{fig : strain_CNN}
\end{figure}

Figs.~\ref{fig : strain} and \ref{fig : strain_CNN} display the scatter plot of the prediction results of the RF and Whisper-based extractors (using, for example, the strain score). In both methods, most predictions robustly approximate the ground truth, and some outliers remain. These outlier significantly contribute to the final RMSE values. Tables~\ref{tab:score correlation ML} and \ref{tab:score correlation NN} list the Pearson correlation values between the CAPE-V scores and prediction results. The correlation value of RF is the same as that of the SOTA, which is the highest among all the approaches. In the case of the method using the pre-trained model, the roughness value shows a low correlation in all pre-trained models. \\

%%% 需要把斜角線畫出來, RF的feature importance要畫一下（既然都說比較好解釋了）

\begin{table}[htbp]
  \centering
  \caption{Correlation value of the CAPE-V score and prediction results using audio parameters as features}
  \label{tab:score correlation ML}
  \begin{tabular}{c c c c c }
    \midrule
    Method & SVM & KNN & RF & SOTA \\
    \midrule
    Severity    & 0.74 & 0.69 & 0.78 & 0.77\\
    Breathiness & 0.57 & 0.54 & 0.64 & 0.75 \\
    Pitch       & 0.64 & 0.60 & 0.65 & 0.64 \\
    Loudness    & 0.61 & 0.58 & 0.70 & 0.71\\
    Roughness   & 0.69 & 0.65 & 0.62 & 0.64 \\
    Strain      & 0.71 & 0.69 & 0.74 & 0.67 \\
    \midrule
    Avg.        & 0.66 & 0.63 & 0.69 &  0.69\\
    \bottomrule
  \end{tabular}
\end{table}

\begin{table}[htbp]
  \centering
  \caption{Correlation value of the CAPE-V score and prediction results using pre-trained models}
  \label{tab:score correlation NN}
  \begin{tabular}{c c c c c c}
    \midrule
    Method   & Wav2vec2 & Hubert & WavLM & Whisper & RF \\
    \midrule
    Severity    & 0.55 & 0.48 & 0.21 & 0.71 & 0.78\\
    Breathiness & 0.65 & 0.69 & 0.41 & 0.71 & 0.64\\
    Pitch       & 0.40 & 0.37 & 0.33 & 0.59 & 0.65\\
    Loudness    & 0.69 & 0.62 & 0.36 & 0.76 & 0.70\\
    Roughness   & 0.16 & 0.24 & 0.18 & 0.28 & 0.62\\
    Strain      & 0.43 & 0.53 & 0.51 & 0.66 & 0.74\\
    \midrule
    Avg.         & 0.48 & 0.49 & 0.33 & 0.62 & 0.69\\
    \bottomrule
  \end{tabular}
\end{table}

\begin{figure}
\centerline{\includegraphics[width=\columnwidth,height=14cm]{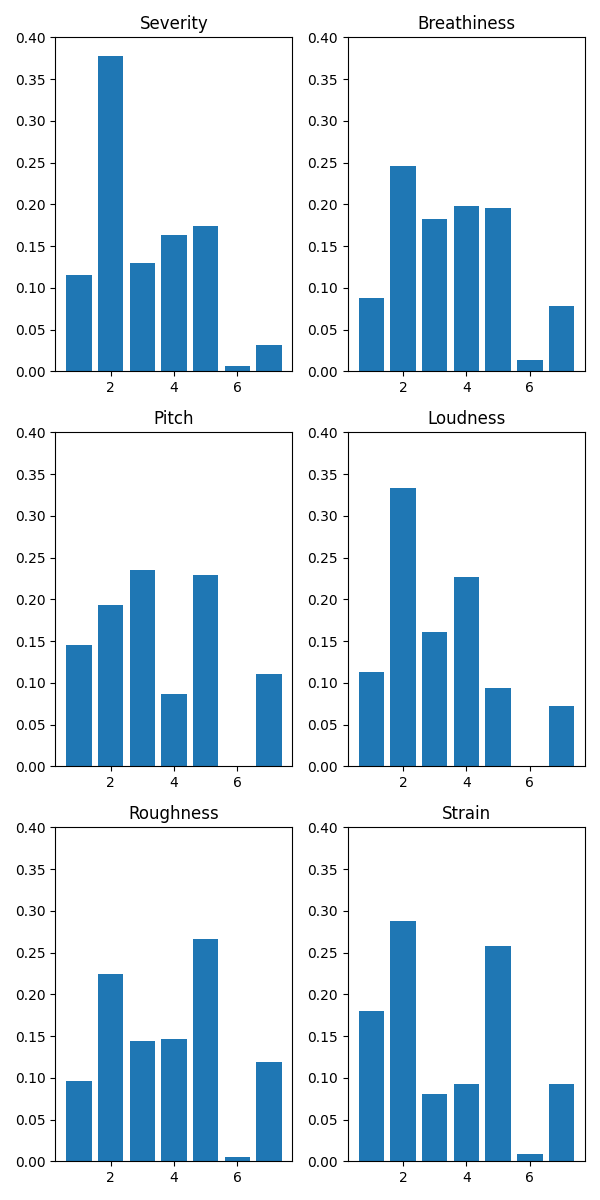}}
\caption{Feature importance of each feature in Random Forest. The numbered features on the X-axis, ranging from 1 to 7, represent zero-crossing, jitter, absolute jitter, shimmer, the HNR, sex, and age, respectively.}
\label{fig : RF}
\end{figure}

Fig.~\ref{fig : RF} depicts the feature importance for RF. The results correspond with those in Table~\ref{tab:correlation}. Jitter, shimmer, and the HNR exhibit greater feature importance than other features. The figures also show that sex was not valued in the RF framework. Combining the information on the correlations and feature importance, we are more convinced that the extracted feature is useful or not. %As for pretrained model extracted representation, \cite{transformer} already reported that audio characteristic can be learned from the transformer, while other information contains in the latent may also be valuable in voice evaluation. \\

\section{Discussion}
In this study, we employed lightly weighted audio parameter extraction and compared it with well-known pre-trained models for feature extraction. Both approaches demonstrated the capacity to extract meaningful features for voice evaluations. RF and Whisper exhibited the RMSE closest to SOTA, whereas RF performed the best in our experiment. 

%% Discussion 1  Attribute that outperform SOTA 
In the RF approach, the RMSE results outperforms those of the SOTA in terms of the roughness and strain attributes. Fig.~\ref{fig : RF} shows that jitter and HNR are the most important features. In \cite{Cape-V}, the roughness attribute and strain were defined based on the irregularity perceived in the voicing source and the perception of excessive vocal effort, respectively. These characteristics can be characterized by both the jitter and HNR. Jitter represents the absence of precise control over the vocal cord, which corresponds with the irregularity in the voicing source, and the HNR denotes the fraction of noise in the emitting sound, which serves as the main characteristic of roughness or strain of sound. Our experimental results proved that the jitter and HNR can be used to characterize the strain and roughness attributes.

%% Discussion 2  Gender information 
Additionally, in our examination, sex was found to be less valued among all features in the RF framework. Although in \cite{Cape-V}, sex and age were also considered in the rating process, the results showed that the sex was the less valued feature for all six scores. Our observations revealed that all features were numeric, whereas sex was encoded by applying one-hot encoding. Numeric features provided more information during prediction, thereby achieving higher feature importance.

%% Discussion 3  Roughness failed in pretrained model.

For the pre-trained model method, no model provided a reliable prediction for the roughness attribute. As stated earlier, the noise component of the voice is important for the perception of voice roughness. However, the implemented pre-trained model was originally designed for ASR task, which relies on noise robustness to achieve improved performance. This decreased the performance in the perception of the roughness attribute of the voice. Pre-existing models have been recognized for their capacity to generate latent representations that augmented predictive abilities for various tasks. Nevertheless, the applicability of pre-existing models has exhibited mixed outcomes \cite{NatureBiomedicalEngineering2022_SSL_review}, particularly when employed within more specialized domains like medicine. Through our experimentation, it has been observed that out-of-domain pre-trained models are capable of producing desired representations that exhibit commendable performance; however, they fall short of surpassing SOTA.

This study had several limitations. Although the augmentation method was employed to overcome the limitations of small datasets, and that noise addition had been proven to enhance robustness in model generalization \cite{vincent2010stacked}, the fine-tuned layers were still easily overfitted and difficult to train. Further investigations into efficient methods for transcending small sample sizes for voice quality assessments are required. In addition, the architecture of the fine-tuned layers of Whisper was different from that of other pre-trained models, which should be considered when comparing their results.  

\section{Conclusion}
In this study, we comprehensively explored voice quality assessment by using different approaches. Audio parameters such as the jitter and HNR were proved to be suitable of characterizing voice quality attributes, particularly roughness and strain. Our findings also showed the usefulness of pre-trained models in extracting features for the evaluation of voices, with defects in noise-related attributes. This study provides insights into employing different feature extraction approachs for voice evaluation, thus laying a foundation for more comprehensive and accurate evaluations.  
%To further optimize the result, limited number of data may be a problem. Although augmentation is carried out, overfitting is still reported in the training phase. 
%Also, scores are given based on the deviance from normal voice. There may be various type of deviance presence, thus increase the difficulty of predictions with limited data. Notably, the ground truth in the experiment is the average value of multiple raters. When contrasting these average values with those provided by individual raters, the resulting RMSE value hovers around 10. This observation implies the presence of a minor inherent error in the initial data collection process.

%overfitting 是操作不良的結果，數值沒有代表性，其實不應該回報

\bibliographystyle{IEEEtran}
\bibliography{refs}

\end{document}